\documentclass{ws-mplb}

\begin{document}

\newcommand{\spd}{$sp$--$d$ }
\newcommand{\ef}{E_{\rm F}}
\newcommand{\du}{{\rm d}}
\newcommand{\e}{{\rm e}}
\newcommand{\Ang}{{\rm \AA}}

\newcommand{\be}{\begin{equation}}
\newcommand{\ee}{\end{equation}}
\newcommand{\ben}{\begin{eqnarray}}
\newcommand{\een}{\end{eqnarray}}
\newcommand{\beq}{\begin{equation}}
\newcommand{\eeq}{\end{equation}}
\newcommand{\B}{\mathrm{B}}
\newcommand{\NB}{\mathrm{NB}}
\newcommand{\RRe}{\mathrm{Re}}
\newcommand{\IIm}{\mathrm{Im}}

\markboth{Trambly de Laissardi\`ere G., Mayou D.}{Quantum Transport in Graphene}

%
\catchline{}{}{}{}{}
%

\title{ELECTRONIC TRANSPORT IN GRAPHENE: QUANTUM EFFECTS AND  ROLE OF LOCAL DEFECTS}

\author{\footnotesize Guy TRAMBLY DE LAISSARDI\`ERE
}

\address{Laboratoire de Physique th\'eorique et Mod\'elisation, CNRS and 
Universit\'e de Cergy-Pontoise, 2 av. A. Chauvin, 95302 Cergy-Pontoise, France\\
guy.trambly@u-cergy.fr}

\author{Didier MAYOU}

\address{Institut N\'eel, CNRS and Universit\'e J. Fourier, BP 166, 38042 Grenoble Cedex 9, France\\
didier.mayou@grenoble.cnrs.fr}

\maketitle

\begin{history}
\end{history}

\begin{abstract}
In this paper we present generic properties of quantum transport in mono-layer graphene. 
In the scheme of the Kubo-Geenwood formula, we compute the square spreading 
of wave packets of a given energy with is directly related to conductivity. 
As a first result, we compute analytically the time dependent diffusion 
for pure graphene. In addition to the semi-classical term a second term exists that is 
due to matrix elements 
of the velocity operator between electron and hole bands. 
This term is related to velocity fluctuations 
i.e. Zitterbewegung effect. 
Secondly, we study numerically  the quantum diffusion in graphene with simple vacancies and pair 
of neighboring vacancies (divacancies), 
that simulate schematically oxidation, hydrogenation and other functionalisations 
of graphene. We analyze in particular the time dependence of the diffusion and its dependence on energy  
in relation with the electronic structure. We compute also the mean free path and the semi-classical 
value of the conductivity as a function of energy in the limit of small concentration of defects.
\end{abstract}


\section{Introduction}

Graphene consists of a mono-layer carbon atoms forming a 2D honeycomb lattice. 
It has attracted much interest because of its unique electronic properties and potential
for device applications.\cite{Novoselov04,Berger04}
Because of the linear dispersion relation of electron states close to Fermi energy, many unconventional
transport behaviors have been shown in graphene. 
The effects of defects and impurities are essential in transport properties.
In particular the role of defect and impurities in a Metal-Insulator transition has been
found experimentally,\cite{Hashimoto04,Wu_X08,Zhou08} but is still not completely understood 
theoretically.$^{4-13}$ 
Atomic vacancy (Va) is one of the natural defects present in graphene  
that has been intensively studied theoretically
(see Ref. \cite{Pereira08a,Pereira08b,Peres10,Wehling10} and Refs. therein).
Moreover, as shown recently by ab initio studies,\cite{Wehling10}
vacancies simulate schematically oxidation and chemical adsorption 
of simple atoms or molecules (H, OH, CH$_3$...).
Indeed, adsorption of atom creates a covalent bond 
between an orbital of this atom and a $p_z$ orbital of one carbon of the graphene sheet.
Therefore, this $p_z$ orbital does not contribute any more to the conduction band ($p_z$ band).
Thus, roughly speaking, the effect on conduction of one adsorption band is simulated
by removing one C atom from graphene sheet.
In the same scheme, an adsorption that creates two covalent bounds on two neighboring 
C atoms  (for instance epoxy group \cite{Incze03}) 
are schematically simulated by a pair of 
two neighboring vacancies (``vacancy pairs'' or divacancies). 

In this paper, we present theoretical investigations of transport properties in graphene. 
In the second part, 
we discuss the link between conductivity and quantum diffusion. 
In the third part,
it is shown that inter-band transitions have significant effect on quantum diffusion in pure graphene. 
In the fourth part, effects of simple vacancies 
and divacancies on quantum diffusion are analyzed.

\section{Conductivity and quantum diffusion}

In the framework of Kubo-Greenwood approach for calculation of the conductivity,
a central quantity is
the average
quadratic spreading of wave packets of energy $E$ at time $t$
along the $x$ direction,
\begin{eqnarray}
  \Delta X^{2}(E,t) =\left\langle \Big(\hat{X}(t)-\hat{X}(0) \Big)^{2}\right\rangle_{E},
 \label{Def_1}
\end{eqnarray}
where $\hat{X}(t)$ it the Heisenberg representation of the position operator $\hat{X}$.
$\langle \hat{A} \rangle_{E}$ means an average of diagonal elements of the operator $\hat{A}$ over all 
states with energy $E$. 
The diffusivity at zero temperature, $\mathcal{D}(E)$, at energy $E$ is deduced from $\Delta X^{2}$,
\begin{eqnarray}
\mathcal{D}(E)  = \lim_{t \rightarrow + \infty} D(E,t) ~~~{\rm with}~~~D(E,t) = \frac{\Delta X^{2}(E,t)}{t},
 \label{Def_2}
\end{eqnarray}
where $D(E,t)$ is called diffusion coefficient. 
In a 2-dimensionnal system with surface $S$, the DC-conductivity $\sigma_{xx}$ at zero temperature 
along the $x$-direction is given by Einstein formula:
\be
\sigma_{xx} (E_{\rm F}) = \frac{e^2}{S} n(E_{\rm F}) \mathcal{D}(E_{\rm F}),
\ee 
where $n(E)$ is the total density of states and $E_{\rm F}$ the Fermi energy.

The effect of decoherence mechanisms such as electron-electron scattering, electron phonon interaction 
(temperature), 
is not considered in the above expression.
This effect can be estimated by introducing an inelastic scattering time $\tau_i$.
$\tau_i$ decreases when the temperature increases. 
In actual graphene at room temperature, realistic values 
of $\tau_i$ is a few $10^{-13}$s.\cite{Wu_X07}
Therefore the experimental conductivity can be estimated by:
\be
\sigma_{xx} (E_{\rm F}) = \frac{e^2}{S} n(E_{\rm F}) \mathcal{D}(E_{\rm F},\tau_i ),
\label{Eq_sigma_tau_i} 
\ee 
The variation of the experimental conductivity with $\tau_i$ is analogous to the variation 
with time which is discussed in detail in this article. 
Here the Fermi-Dirac distribution function is taken equal to its zero temperature
value. This is valid provided that the electronic properties vary smoothly on the
thermal energy scale $k_{\rm B}T$.

\section{Quantum diffusion in pure graphene}

We have developed\cite{NanoLet10} a simple tight binding (TB) scheme that reproduces 
the {ab initio} calculations 
of the electronic states with energies $\pm 2$\,eV around the energy of Dirac point ($E=0$).
Only $p_z$ orbitals are taken into account since we are interested in what happens at the Fermi level. 
The Hamiltonian is,
\ben
\hat{H} &=&  \sum_{<i,j>} t_{ij} \, |i\rangle \langle j| ,
\label{Eq_hamilt}
\een
where coupling matrix element, $t_{ij}=\langle i|\hat{H}|j\rangle$, 
depends on the distance $r_{ij}$ between orbitals $|i\rangle$ and $|j\rangle$,
\ben
t_{ij} ~=~ -\gamma_0 \,{\rm e}^{q_{\pi}    \left(1-{r_{ij}}/{a} \right)}, 
\label{eq:tb0}
\een
with $a=1.418$\,{\rm \AA}, the firt neighbor distance.
First neighbors interaction in
a plane is taken equal to $\gamma_0 = 2.7$\,eV.
Second neighbors interaction $\gamma_0'$ 
in a plane equals\cite{CastroNeto09} to
$0.1\times\gamma_0$ 
which fixes value of the $q_{\pi}$ in (\ref{eq:tb0}).

In crystals, once the band structure is calculated from the tight-binding Hamiltonian
the average quadratic spreading
can be computed exactly in the basis
of Bloch states.\cite{PRL06}
In pure graphene the average square spreading is the sum of two terms:\cite{PRL06}
\begin{equation}
\Delta X^2(E,t) = V_{\B}^2 t^2 + \Delta X_{\NB}^2(E,t).
\end{equation}
The first term is the ballistic (intra-band)  contribution
at the energy $E$.
$V_B$ is the Boltzmann velocity in the $x$ direction. 
The semi-classical theory
is equivalent to taking into account only this first term.
The second term (inter-band contributions),
$\Delta X^2_{\NB}(E,t)$,  is
a non-ballistic (non-Boltzmann) contribution.
It is
due to the non-diagonal
elements in the eigenstates basis $\{| n\rangle \}$  of the velocity operator $\hat{V}_x$, 
\begin{eqnarray}
\hat{V}_x = \frac{1}{i \hbar}~ \Big[ \hat{X} , \hat{H} \Big].
\end{eqnarray}
From the definition (\ref{Def_1}), one obtains,\cite{PRL06,Mayou08RevueTransp}
\begin{equation}
\Delta X_{\NB}^2(E,t) = 2 \hbar^2
\left\langle
\sum_{\vec k, n' (n' \ne n)}
\left[ 1 - \cos\left(\frac{(E_{\vec k,n}-E_{\vec k,n'})t}{\hbar} \right)\right] 
\frac{\left| \langle n\vec k | \hat{V}_x | n'\vec k \rangle \right|^2}{(E_{\vec k,n}-E_{\vec k,n'})^2}
\right\rangle_{E_{\vec k,n}=E} 
\label{Eq_DeltaX2_NB}
\end{equation}
where $E_{\vec k,n}$  is energies of eigenstate $| n \vec k \rangle$.

In graphene, assuming a restriction of the Hamiltonian (\ref{Eq_hamilt}) 
to the first neighbor interactions only, 
$\Delta X_{\NB}^2(E,t)$ is given by :
\begin{equation}
\Delta X_{\NB}^2(E,t) = \frac{V_B^2\hbar^2}{2 E^2} \left( 1 - 
\cos \frac{2E}{\hbar}t   \right) ~~~{\rm with}~~V_{_\B} = \frac{3 a \gamma_0}{2\sqrt{2} \hbar}.
\label{Eq_DeltaX2_NB_graphene}
\end{equation}
At small time, $t < \hbar/E$,  the Non-Boltzmann term is equal to Boltzmann 
term $\Delta X_{\NB}^2(E,t) \simeq V_{_\B}^2t^2$, 
thus $\Delta X^2(E,t) \simeq 2 V_{_\B}^2t^2$ and $D(E,t) \simeq 2V_{_\B}^2t$ . 
Whereas for large $t$, the Boltzmann term dominates and  
$\Delta X^2(E,t) \simeq V_{_\B}^2t^2$ and $D(E,t) \simeq V_{_\B}^2t$.
The non-Boltzmann term is due to matrix elements of the velocity operator 
between the two bands (i.e. between the hole and electron states having the same wavevector). 
These matrix elements imply that the velocity correlation function has also two parts: 
one constant and the other oscillating at a frequency $2 E/ \hbar $ where E is the energy of the state. 
This is precisely the phenomenon of jittery motion also called Zitterbewegung. 
Note that in any crystal having several bands there are also components of 
the velocity correlation function which are oscillating at frequencies 
$(E_{\vec k,n}-E_{\vec k,n'})/ \hbar $. 
Therefore Zitterbewegung is quite common in condensed matter physics. 
For example approximants of quasicrystals present very strong Zitterbewegung effect and 
the non-Boltzmann contribution dominates the Boltzmann 
contribution.\cite{PRL06,Triozon02,Trambly10b}

\section{Quantum diffusion in the presence  of vacancies}

\begin{figure}[t]
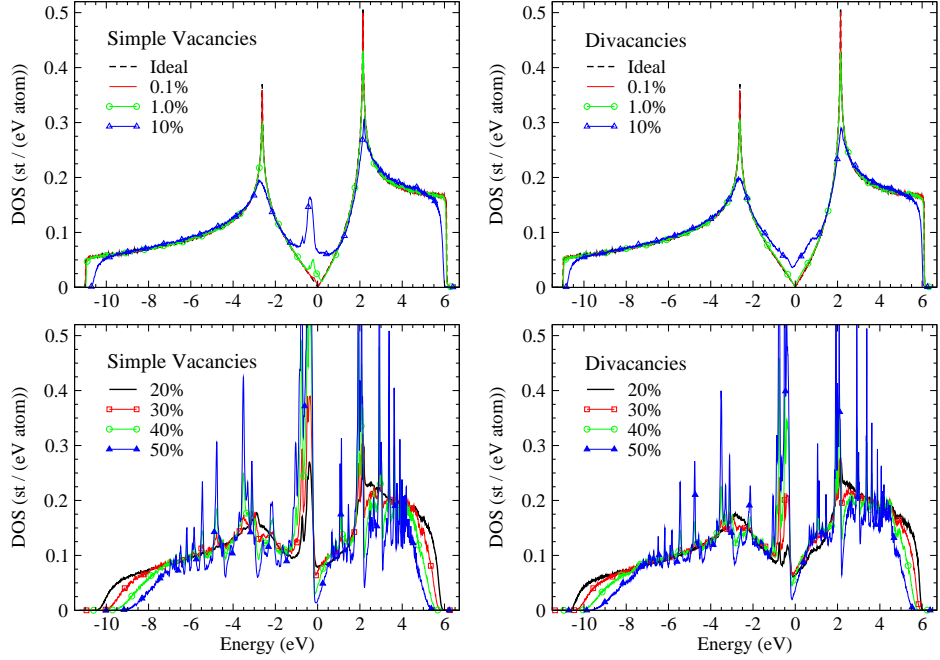

\centerline{\psfig{file=FigMPLB_DOS_0-10La.eps,width=6cm}~~~\psfig{file=FigMPLB_DOS_0-10paireLa.eps,width=6cm}}
\vspace*{4pt}

\centerline{\psfig{file=FigMPLB_DOS_20-50La.eps,width=6cm}~~~\psfig{file=FigMPLB_DOS_20-50paireLa.eps,width=6cm}}
\vspace*{8pt}
\caption{
DOS in graphene for various concentration of simple vacancies and divacancies.
\label{Fig_DOS}}
\end{figure}

In this part, the density of states (DOS) and diffusion coefficient calculated numerically 
in disorder graphene are presented.
We study both the effects of simple vacancies and the effects of pairs of nearest neighbor vacancies
(divacancies).
The total DOS is computed by a Lanczos-type recursion method. 
In the scheme of the Kubo-Greenwood formula for transport properties, 
the diffusion coefficient, $D(E,t)$, is computed by using 
the PEQD method (Polynomial Expansion for Quantum Diffusion method) 
developped by the Mayou, Khanna, Roche and Triozon.\cite{Mayou88,Mayou95,Roche97,Triozon02}
This method 
allows very efficient numerical calculations by recursion in real-space.
This method has been  used to study quantum transport 
in disordered graphene\cite{Lherbier08} and 
chemically doped graphene,\cite{Lherbier08b} 
graphene with Ozone functionalization,\cite{Leconte10}
graphene with structural defects (Stone-Wales and divacancies)\cite{Lherbier10}.
We use periodic boundary conditions with a supercell size $L_x = 2947.2$\,nm, $L_y = 122.8$\,nm.
The energy resolution is about $0.01$\,eV.
Defects (simple vacancies and divacancies) are randomly distributed in a supercell.
We have checked that the present results do not depend on the size of the supercell.

Electronic structure of simple vacancies has been intensively 
studied.\cite{Pereira06,Pereira08a,Pereira08b,Peres06,Wu_S08} 
Graphene lattice is a bipartite lattice with two 
equivalent sublattices $A$ and $B$. 
If we consider an Hamiltonian containing only nearest-neighbor hopping
(uncompensated lattice\cite{Pereira06,Pereira08a}) 
each $A$ atom is coupled with only $B$ atoms (and reciprocally). 
Consequently, at very low concentration of vacancies ($\sim$ isolated simple vacancies), 
the distribution of vacancies is locally uneven between the two 
sublattices and zero energy states necessarily appears.\cite{Brouwer02,Pereira08a}
But if vacancies are arranged by pair of nearest neighbor vacancies (divacancies), 
the electronic structure at low impurities concentration is completely different. 
Indeed in that case, the  distribution of vacancies is locally even between the two 
sublattices, and  zero energy state does not occur.

\begin{figure}[t]
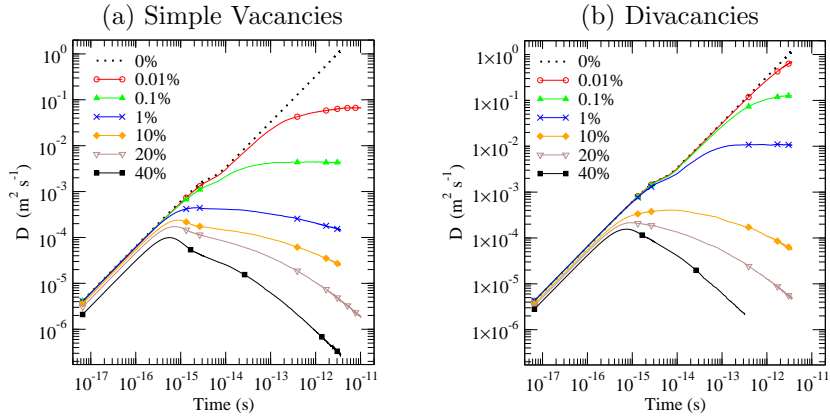

~~~~~~~~~~~~~~~~~~(a) Simple Vacancies~~~~~~~~~~~~~~~~~~~~~~~~~~~(b) Divacancies
\vspace*{2pt}

\centerline{\psfig{file=FigMPLB_La_DX2-sT_E-_23.eps,height=5cm}~~~~~~~\psfig{file=FigMPLB_pLa_DX2-sT_E-_23.eps,height=5cm}}
\vspace*{8pt}
\caption{
Diffusion coefficient, $ D={\Delta X^{2}}/{t}$, versus time $t$ for wave packets at energy 
$E$ in the peak of simple vacancies: 
$E=-0.23$\,eV.
\label{Fig_DX2-sT_E-.23}}
\end{figure}

\begin{figure}[]
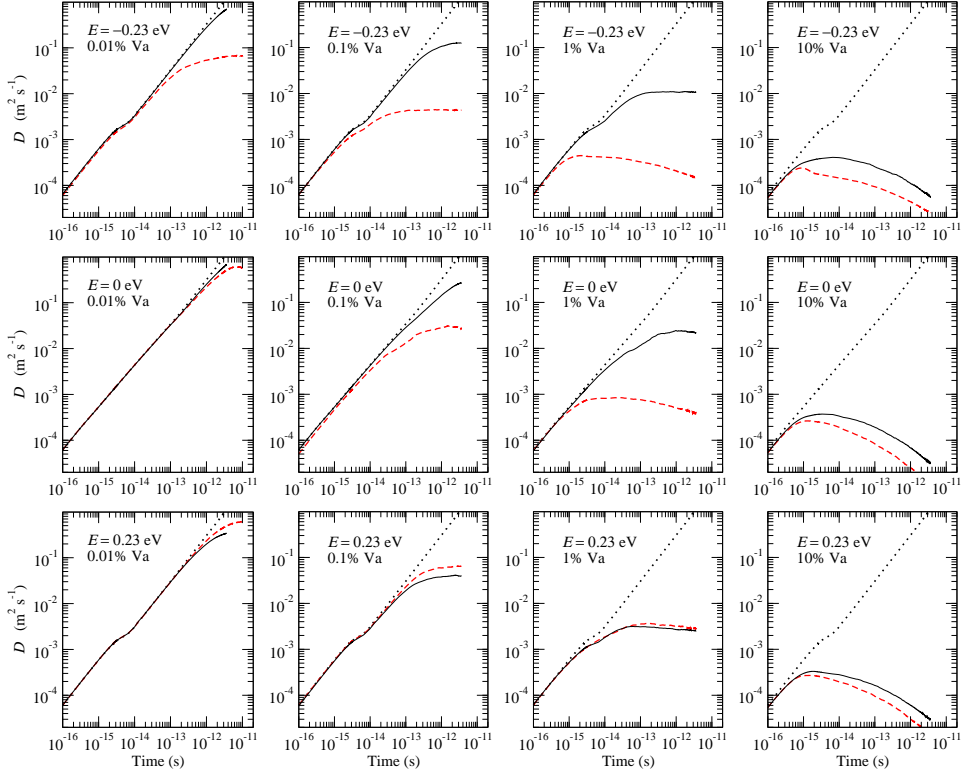

\centerline{\psfig{file=FigMPLB_0_01La+pLa_DX2sT_E-_23.eps,width=3.35cm}
\psfig{file=FigMPLB_0_1La+pLa_DX2sT_E-_23.eps,width=3cm}
\psfig{file=FigMPLB_1La+pLa_DX2sT_E-_23.eps,width=3cm}
\psfig{file=FigMPLB_10La+pLa_DX2sT_E-_23.eps,width=3cm}}
\vspace*{4pt}

\centerline{\psfig{file=FigMPLB_0_01La+pLa_DX2sT_E0.eps,width=3.35cm}
\psfig{file=FigMPLB_0_1La+pLa_DX2sT_E0.eps,width=3cm}
\psfig{file=FigMPLB_1La+pLa_DX2sT_E0.eps,width=3cm}
\psfig{file=FigMPLB_10La+pLa_DX2sT_E0.eps,width=3cm}}
\vspace*{4pt}

\centerline{\psfig{file=FigMPLB_0_01La+pLa_DX2sT_E_23.eps,width=3.35cm}
\psfig{file=FigMPLB_0_1La+pLa_DX2sT_E_23.eps,width=3cm}
\psfig{file=FigMPLB_1La+pLa_DX2sT_E_23.eps,width=3cm}
\psfig{file=FigMPLB_10La+pLa_DX2sT_E_23.eps,width=3cm}}
\vspace*{8pt}
\caption{
Diffusion coefficient, $D={\Delta X^{2}}/{t}$, versus time $t$ for wave packets at energy $E$, 
with various concentrations of vacancies: (point line) perfect graphene, (full line) divacancies, 
(dashed line) simple vacancies. 
\label{Fig_DX2-st_E_La}}
\end{figure}

A finite concentration $c$ of simple vacancies induced a small enlargement of the zero energy 
modes.\cite{Pereira06,Wu_S08}
Moreover, for Hamiltonian with second neighbor hopping, these states are strongly 
enlarged and their energy is shifted to negative values.\cite{Pereira06,Wu_S08}
As shown figure \ref{Fig_DOS}, our TB model reproduces this behavior in good  agreement with 
previous studies. The states due to simple vacancies is observed at the energy $E_p \simeq -0.23$\,eV.
As expected, the peak at   $E_p \simeq -0.23$\,eV is not obtained for small concentration of divacancies
(figure \ref{Fig_DOS}). For larger concentration of divacancies the distribution 
of vacancies could be locally uneven between the two sublattices and the peak at  $E_p \simeq -0.23$\,eV
is found. 
For large concentration of simple vacancies and divacancies, $c > 30\%$,
a low density of states is obtained around $E=0$ and gaps may occur for $c \ge 50\%$.

The diffusive coefficient $D$, $D=\Delta X^2 / t$,
versus time $t$, is shown on figure \ref{Fig_DX2-sT_E-.23} 
for an energy in the peak of simple vacancies, $E=-0.23$.  
Different regimes are observed. 
For small time $t$, the effect of elastic scattering is very small and the regime is ballistic, 
$D(E,t) = V^2 t$. 
As explain in the previous section, the effective velocity is $V = \sqrt{2} V_B$ for $t<\hbar/E$ 
and $V = V_B$ for $t>\hbar/E$, which leads, in this log-log scale, to a shift  of $D(t)$
 at $t \simeq \hbar/E$ 
(figures \ref{Fig_DX2-sT_E-.23} and \ref{Fig_DX2-st_E_La}).
After the ballistic regime  $D$ reaches  a maximum values equals to the semi-classical 
diffusivity $\mathcal{D}_{sc}(E)$. 
Indeed if there were no localization effects the diffusivity would be constant at larger times.
The elastic scattering time $\tau_e(E)$ is defined by $\tau_e = \mathcal{D}_{sc}/V_{\rm B}^2$,
and
the elastic mean free path, $l_e$, is deduced from
$l_e = V_{\rm B} \tau_e$. 
At large time, in 2D-system with defects, quantum interferences lead to decay of $D(t)$ 
with a characteristic localization length $\xi$.\cite{Lee85}
When $l_e \ll \xi$, the semi-classical diffusive regime (saturation of $D$, $D(E,t) \simeq D_{sc}(E)$) 
occurs on a large time scale.
But for $l_e \simeq \xi$, the semi-classical approximation breaks down and the decay of $D(t)$ 
is strong for $t>\tau_e$.

\begin{figure}[]
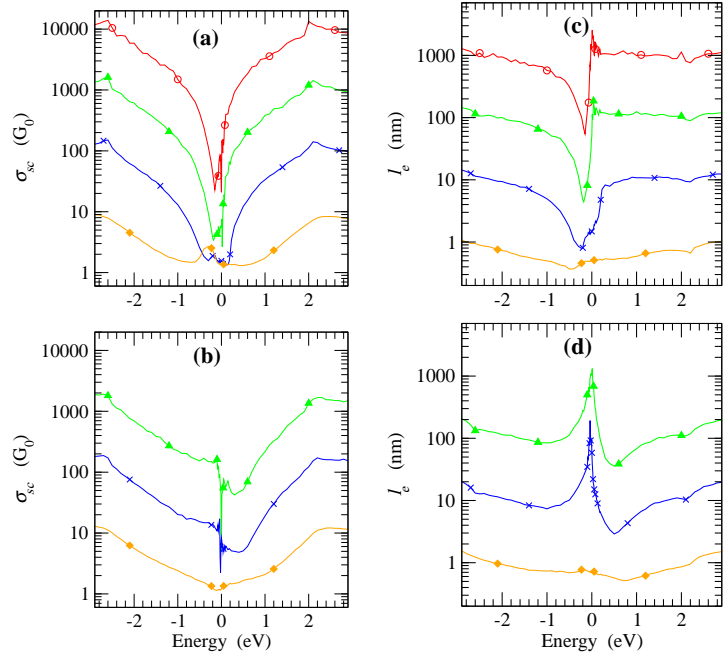

\centerline{\psfig{file=FigMPLB_Sigma_sc_La_v2.eps,width=4.5cm}~~~~\psfig{file=FigMPLB_l_e_La_v2.eps,width=4.5cm}}
\vspace*{4pt}

\centerline{\psfig{file=FigMPLB_Sigma_sc_paireLa_v2.eps,width=4.5cm}~~~~\psfig{file=FigMPLB_l_e_paireLa_v2.eps,width=4.5cm}}
\vspace*{8pt}
\caption{
Semi-classical conductivity $\sigma_{sc}$ (in unit of $G_0 = {2 e^2}/{h}$) 
and elastic mean free path $l_e$ versus energy $E$:
(a) and (c) simple vacancies, (b) and (d) divacancies. 
Concentration of defects:
cercle $c=0.01$\%,  triangle $c=0.1$\%, cross $c=1$\%, diamond $c=10$\%.
\label{Fig_Sigma_Le}}
\end{figure}

The mean free path and semi-classical value of the conductivity are shown in figure  
\ref{Fig_Sigma_Le} as a function of energy and for different concentrations. 
The semi-classical value of the conductivity is  
$\sigma_{sc} (E_{\rm F}) = \frac{e^2}{S} n(E_{\rm F}) \mathcal{D}_{sc} (E_{\rm F})$,
where  $\mathcal{D}_{sc} (E_{\rm F}) $ is  semi-classical diffusivity defined above.
Let us recall that when the semi-classical of conductivity is  much larger than the quantum 
of conductance, $G_0 = {2 e^2}/{h}$, the localization length is much larger than the mean free 
path and there is a large scale of inelastic scattering times ${\tau_i}$   
on which the quantum interferences are small and the  semi-classical theory of transport is valid.

One sees from figure \ref{Fig_Sigma_Le} that for sufficiently low doping the conductivity and mean 
free path are inversely proportional to the concentration of defects in agreement with the 
Fermi Golden Rule. For divacancies the mean free path increases when the energy tends to zero. 
This is in agreement with the semi-classical model since it can be shwon that the T-matrix 
of the divacancy is energy independent close to the Dirac point and therefore th mean free 
path is inversely proportional to the energy. 
For simple vacancy,
it is also clear that close to the energy of the resonant state $E=-0.23$\,eV
the scattering is strong and accordingly the semi-classical conductivity and mean free path are small.
In the energy region $-0.4<E<0.4$\,eV which is of most interest for experiments the mean free path 
can vary by a factor of about ten. This variation is partly compensated by the variation of the density 
of states so that the semi-classical conductivity varies by a factor of about 3 at most in 
the same energy range.

\begin{figure}[]
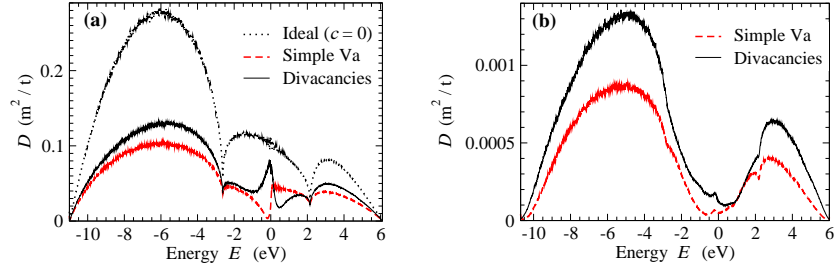

\centerline{\psfig{file=FigMPLB_D_La.1pc_t_3_26E-13.eps,height=3.5cm}~~~~~\psfig{file=FigMPLB_D_La10pc_t_3_26E-13.eps,height=3.5cm}}
\vspace*{8pt}
\caption{
Diffusion coefficient, $ D={\Delta X^{2}}/{t}$, versus energy $E$ of wave packets at time 
$t=3.26\times 10^{-13}$s, 
for (a) $c=0.1$\% vacancies, (b) $c=10$\% vacancies.
\label{Fig_DX2-sT_t3.26E-13}}
\end{figure}

The diffusive coefficient $D$, $D=\Delta X^2 / t$, versus the energy $E$ of the wave packets
is shown on figure  \ref{Fig_DX2-sT_t3.26E-13} for two concentrations of defects.
Figures \ref{Fig_DX2-st_E_La} and \ref{Fig_DX2-sT_t3.26E-13} compare diffusion coefficient 
$D$ in both case: 
simple vacancies and divacancies. 
For energies out of the peak in the DOS due to simple vacancies ($E=0$ and $E=0.23$\,eV 
in figure \ref{Fig_DX2-st_E_La}), 
this two kinds of defects have similar consequence on 
diffusion. 
But for energy in the peak ($E \simeq - 0.23$\,eV) and defect concentration $c<10$\%, 
electron localization by simple vacancies is stronger.  
This is due to resonant scattering by the impurity potential. 
This is very similar to scattering by virtual bound states in the physics 
of transition metals for example.
As shown figure \ref{Fig_DX2-sT_t3.26E-13}(b), for large concentration of defects 
($c \ge 10$\%),  
simple vacancies and divacancies have similar scattering effects.

\section{Conclusion}

To conclude we have studied in detail the quantum diffusion of electrons 
in graphene in the presence of local defects. 
These local defects are simple vacancies and divacancies. 
They are representative of functionnalized graphene with molecule that produce covalent bonding either 
with one carbon atom (simple vacancies represent for example fonctionalization by Hydrogen)) 
or with two neighboring  carbon atoms 
(divacancies can represent functionalization by one oxygen atom (epoxy group)). 
The time dependence reveals 
different regimes:  ballistic with a kink at short time, 
saturation, existence of a plateau if  the localization length is much 
greater than the mean free path, decrease of the diffusivity due to localization. 
We show also important differencies  between the electronic structure of simple vacancies 
and pairs of nearest neighbor vacancies (divacancies). 
In the case of simple vacancies a narrow maximum 
occurs in the density of states due to the existence of an impurity resonance. 
Just at the impurity energy ($E \simeq- 0.23$\,eV) the scattering is strong and the 
tendency to localization is enhanced. 
Finally, we present also the mean free path and the 
semi-classical value of the conductivity as a function of energy in the limit of small 
concentration of defects. Let us recall that when the semi-classical value of 
the conductivity is much larger than the quantum of conductance ${2 e^2}/{h}$ 
the localization length is much larger than the mean free path and there is a large 
scale of inelastic scattering times ${\tau_i}$ on which the quantum interferences 
are small and the  semi-classical theory of transport is valid.

\section*{Acknowledgments}
We thank
L. Magaud
for fruitful discussions.
The computations have been performed at the
Service Informatique Recherche (SIR),
Universit\'e de Cergy-Pontoise.
We thank Y. Costes and David Domergue, SIR,
for computing assistance.

\appendix


\begin{thebibliography}{0}

\bibitem{Novoselov04} K. S. Novoselov, A. K. Geim, S. V. Morozov, D. Jiang, Y. Zhang, 
S. V. Dubonos, I . V. Grigorieva and A. A. Firsov,
{\it Science}, {\bf 306} (2004) 666. 
 
\bibitem{Berger04} C. Berger, Z. M. Song, T. B. Li, X. B. Li, A. Y. Ogbazghi, 
R. Feng, Z. T. Dai, A. N. Marchenkov, E. H. Conrad, P. N. First and W. A. de Heer, 
{\it J Phys. Chem. B} {\bf 108} (2004) 19912. 

\bibitem{Hashimoto04} A. Hashimoto, K. Suenaga, A. Gloter, K. Urita 
and S. Iijima, 
{\it Nature} {\bf 430} (2004) 870. 

\bibitem{Wu_X08} X. Wu, M. prinkle, X. Li, F. Ming, C. Berger and W. A. de Heer, 
{\it Phys. Rev. Lett.} {\bf 101} (2008) 026801. 

\bibitem{Zhou08} S. Y. Zhou, D. A. Siegel, A. V. Fedorov, and A. Lanzara,
{\it Phys. Rev. Lett.} {\bf 101} (2008) 086402. 

\bibitem{Pereira06} V. M. Pereira, F. Guinea, J. M. B. Lopes dos Santos, 
N. M. R. Peres and A. H. Castro Neto. 
{\it Phys. Rev. Lett.} {\bf 96} (2006) 036801. 

\bibitem{Pereira08a} V. M. Pereira, J. M. B. Lopes dos Santos 
and A. H. Castro Neto, 
{\it Phys. Rev. B} {\bf 77} (2008) 115109. 

\bibitem{Pereira08b} A. L. C. Pereira and P. A. Schulz, 
{\it Phys. Rev. B} {\bf 78} (2008) 125402. 

\bibitem{Lherbier08} A. Lherbier, B. Biel, Y.-M. Niquet and S. Roche, 
{\it Phys. Rev. Lett.} {\bf 100} (2008) 036803.

\bibitem{Lherbier08b} A. Lherbier, X. Blase, Y.-M. Niquet, F. Triozon and S. Roche. 
{\it Phys. Rev. Lett.} {\bf 101} (2008) 036808. 

\bibitem{Leconte10} N. Leconte, J. Moser, P. Ordejon, H. Tao, 
A. Lherbier, A. Bachtold, F.c Alsina, 
C. M. Sotomayor Torres, J.-C. Charlier and S. Roche, 
{\it ACS Nano} {\bf 4} (2010) 4033. 

\bibitem{Lherbier10} A. Lherbier, S. M.-M. Dubois, X. Declerck, 
S. Roche, Y.-M. Niquet and J.-C. Charlier, 
{\it Cond-Mater. arXiv:1012.4955}  (2010). 

\bibitem{Peres10} N. M. R. Peres,
{\it Journal of Physics: Condensed Matter} 
{\bf 21} (2009) 323201.

\bibitem{Wehling10} 
T. O. Wehling, S. Yuan, A. I. Lichtenstein,  A. K. Geim, and M. I. Katsnelson,
{\it Phys. Rev. Lett.} 
{\bf 105} (2010) 056802.


\bibitem{Incze03} A. Incze, A. Pasturel and C. Chatillon,
{\it Surface Science} 
{\bf 537} (2003) 55. 

\bibitem{Wu_X07} X. Wu, X. Li, Z. Song, C. Berger 
and W. A. de Heer,
{\it Phys. Rev. Lett.} {\bf 98} (2007) 136801. 

\bibitem{NanoLet10} G. Trambly de laissardi\`ere, D. Mayou and L. Magaud,
{\it Nano Lett.} {\bf 10} (2010) 804. 

\bibitem{CastroNeto09} A. H. Castro Neto, F. Guinea, N. M. R. Peres, K. S. Novoselov 
and A. K. Geim,
{\it Rev. Mod. Phys.} {\bf 81} (2009) 109. 

\bibitem{PRL06} G. Trambly de Laissardi\`ere, J.-P. Julien and D. Mayou,
{\it Phys. Rev. Lett.} {\bf 97} (2006) 026601. 

\bibitem{Mayou08RevueTransp} D. Mayou and G. Trambly de Laissardi\`ere,
in Quasicrystlas
Ed. T. Fujiwara and Y. Ishii, 
Handbook of Metal Physics, Vol. 3,
(Elsevier, 2008) p. 209. 

\bibitem{Triozon02} F. Triozon, Julien Vidal, R. Mosseri, and Didier Mayou,
{\it Phys. Rev. B} {\bf 65} (2002) 220202. 

\bibitem{Trambly10b} G. Trambly de Laissardi\`ere, C. Oguey and D. Mayou,
{\it Philosophical Magazine} (2011) in press. 

\bibitem{Mayou88} D. Mayou,
{\it Europhys. Lett.} {\bf 6} (1988) 549. 

\bibitem{Mayou95} D. Mayou and S.N. Khanna, 
{\it J. Phys. I Paris} {\bf 5} (1995) 1199.

\bibitem{Roche97} S. Roche and D. Mayou, 
{\it Phys. Rev. Lett.} {\bf 79} (1997) 2518,

\bibitem{Peres06} N. M. R. Peres, F. Guinea and A. H. Castro Neto, 
{\it Phys. Rev. B} {\bf 73} (2006) 125411, 

\bibitem{Wu_S08} S. Wu, Lei Jing, Q. Li, Q. W. Shi, J. Chen, H. Su, X. Wang and J. Yang, 
{\it Phys. Rev. B} {\bf 77} (2008) 195411, 

\bibitem{Brouwer02} P. W. Brouwer, E. Racine, A. Furusaki, Y. Hatsugai, Y. Morita and C. Mudry,
{\it Phys. Rev. B} {\bf 66} (2002) 014204. 

\bibitem{Lee85} P. A. Lee and T. V. Ramakrishnan,
{\it Rev. Mod. Phys.} {\bf 57} (1985) 287.


\end{thebibliography}

%
%
%
%
%

\end{document}